\documentclass[showpacs,aps,amsmath,amssymb,preprint,preprintnumbers]{revtex4}
\newcommand{\BE}{\begin{equation}}
\newcommand{\EE}{\end{equation}}
\newcommand{\BA}{\begin{eqnarray}}
\newcommand{\EA}{\end{eqnarray}}
\usepackage{graphicx}
\usepackage{dcolumn}
\usepackage{bm}
\input epsf.tex

\begin{document}

\title{Left-Handed-Material-like behavior revealed by arrays of dielectric cylinders}

\author{Chao-Hsien Kuo}\author{Zhen Ye}\email{zhen@phy.ncu.edu.tw}
\affiliation{Wave Phenomena Laboratory, Department of Physics,
National Central University, Chungli, Taiwan}

\date{October 22, 2003}

\begin{abstract}

We investigate the electromagnetic propagation in two-dimensional
photonic crystals, formed by parallel dielectric cylinders
embedded a uniform medium. The transmission of electromagnetic
waves through prism structures are calculated by the standard
multiple scattering theory. The results demonstrate that in
certain frequency regimes and when the propagation inside the
scattering media is not considered, the transmission behavior
mimics that expected for a left-handed material. Such feature may
illusively lead to the conclusion that a left-handed material is
fabricated and it obeys Snell's law of negative refraction. We
also discuss possible ambiguities that may be involved in previous
experimental evidence of left-handed materials.

\end{abstract}

 \maketitle

The concept of the so called Left Handed Material (LHM) or
Negative Refraction Index Material (NRIM) was first proposed by
Veselago many years ago \cite{Ves}. In the year 2000, Pendry
proposed that a lens made of LHM can overcome the traditional
limitation on the optical resolution and therefore make ``perfect"
images \cite{Pendry}. Since then, the search for LHM and the study
of the properties of LHM have been skyrocketing, signified by the
rapid growth of the related literatures.

An earlier realization of LHM consisted of composite resonator
structures of metallic wires and rings operated in the regime of
microwaves \cite{exp}. The measurements involved refraction of
microwaves by a prism. As pointed out in \cite{Chuang}, however,
while this work is revolutionary, there are a few unaddressed
questions. These include problems associated with such as near
field effects due to either rapid dispersion along the propagation
direction \cite{Valanju} or smaller amount of attenuation on the
shorter side of the prism \cite{Garcia}. Moreover, the experiment
had only measured a single angled prism; this is sufficient only
if the {\it a priori} assumption that the material is refractive
is valid. These shortcomings allow for alternative interpretation
of the reported experimental data, as discussed in
Ref.~\cite{Valanju,Garcia}.

Ever since its inception, theoreticians have questioned the
concept of LHM and the perfect lenses made of LHM. Such a
challenge can be classified into two categories. First, scientists
have questioned whether left-handed materials are genuinely
possible. Second, even if they existed, whether LHM would really
make perfect lenses is also debatable \cite{Efros,Hooft,comm,Ye}.
The principle behind the first inquiry is that although peculiar
phenomena observed for some artificial materials may be in
conflict with our immediate intuition, if they can be still
explained in the framework of current knowledge, the resort to
negative refraction or LHM is at least not a definite necessity.

Recently, new sets of experimental measurements have been reported
to provide affirmative evidences on LHM. For instance, Houck et
al. \cite{Chuang} measured two dimensional profiles of collimated
microwave beams transmitted through composite wire and split-ring
resonator prisms. The authors used two angled prisms, in an
attempt to refute alternatives posed in the criticisms on earlier
measurements, to obtain a rather consistent negative refractive
index. The data appeared to obey Snell's law. The experimental
setup in Ref.~\cite{Chuang} represents a common method of deducing
the refractive index of LHM. Based upon our rigorous simulations,
however, we believe that these experimental results are not
conclusive. They may not be sufficient to confirm the existence of
LHM. As a matter of fact, we found that the apparent abnormal
refraction that has been thought to be the negative refraction may
be well explained in the context of the partial-gap effects or the
related Bragg scattering, which are common for all kinds of wave
propagation in periodic structures. In this Letter, we present
some key simulation results to support our claim.

Here we consider the transmission of electromagnetic (EM) waves in
photonic crystals. To make it easy to reproduce our results, we
will use the photonic crystal structures that have been commonly
used in previous simulations, such as those in \cite{PRB}. The
systems are two dimensional photonic crystals made of arrays of
parallel dielectric cylinders placed in a unform medium, which we
assume to be air. The solution for the wave scattering and
propagation in such systems can be obtained by the multiple
scattering theory. This theory is exact and was first formulated
systematically by Twersky \cite{Twersky}, then has been
reformulated and applied successfully to optical, sonic and water
wave problems \cite{JOSA,Chen,Yue}. The results show that an
apparent `negative' refraction is indeed possible, in line with
the experimental observation \cite{Chuang}. But, such a negative
refraction is not sufficient in proving that the materials are
left handed.

The multiple scattering theory (MST) used in our simulation can be
summarized as follows. Consider an arbitrary array of dielectric
cylinders in a uniform medium. The cylinders are impinged by an
optical source. In response to the incident wave from the
transmitting source and the scattered waves from other scatterers,
each scatterer will scatter repeatedly waves and thus scattered
waves can be expressed in terms of a modal series of partial
waves. Regarding these scattered waves as the incident wave to
other scatterers, a set of coupled equations can be formulated and
computed rigorously. The total wave at any spatial point is the
summation of the direct wave from the source and the scattered
waves from all scatterers. The intensity of the waves is
represented by the modulus of the wave field. The details about
MST can be found in Ref.~\cite{JAP}.

For brevity, we only consider the E-polarized waves, that is, the
electric field is kept parallel to the cylinders. The following
parameters are used in the simulation. (1) The dielectric constant
of the cylinders is 14, and the cylinders are arranged in air to
form a square lattice. (2) The lattice constant is $a$ and the
radius of the cylinders is 0.3$a$; in the computation, all lengths
are scaled by the lattice constant, so all the lengths are
dimensionless. (3) A variety of the outer shapes of the arrays is
considered, including the prism structures and the slab.

First we consider the propagation of EM waves through the prism
structures of arrays of dielectric cylinders, by analogy with
those shown in \cite{Chuang}. Two sizes of the prisms are
considered and illustrated by Fig.~\ref{fig1}. We have used two
types of sources in the simulations: (1) plane waves; (2)
collimated waves by guiding the wave propagation through a window
before incidence on the prisms, mimicking most experiments; (3) a
line source. The results from these three scenarios are similar.
In this Letter, we only show the results from the line source
which is located at some distance beneath the prisms. We note here
that when using a plane wave, effects from the prism edges play a
role and should be removed.

We have plotted the transmitted intensity fields. The results are
presented in Fig.~\ref{fig1}. Here the incident waves are
transmitted vertically from the bottom. The impinging frequency is
0.192$*2\pi c/a$; the frequency has been scaled to be
non-dimensional in the same way as in Ref.~\cite{PRB}. We note
that at this frequency, the wave length is about five times of the
lattice constant, nearly the same as that used in the experimental
of \cite{Chuang}. The incidence is along the [$\cos 22.5^o$, $\sin
67.5^o$] direction, that is, the incidence makes an equal angle of
22.5$^o$ with regard to the [10] and [11] directions of the square
lattice; the reason why we choose this direction will become clear
from later discussions. Moreover, on purpose, the intensity
imaging is plotted for the fields inside and outside the prisms on
separate graphs which have been scripted by `1' and `2'
respectively.

Without looking at the fields inside the prisms, purely from
Fig.~\ref{fig1} (a1) and (b1) we are able to calculate the main
paths of the transmitted intensities. The geometries of the
transmission are indicated in the diagrams. The tilt angles of the
prisms are denoted by $\phi$, whereas the angles made by the
outgoing intensities relative to the normals of the titled
interfaces are represented by $\theta$. According to the
prescription outlined in Ref.~\cite{Chuang}, once $\phi$ and
$\theta$ are determined, Snell's law is applied to determine the
effective refractive index: $n\sin\phi = \sin\theta$. If the angle
$\theta$ is towards the higher side of the prisms with reference
to the normal, the angle is considered positive. Otherwise, it is
regarded as negative. In light of these considerations, the
apparent `negative refractions', similar to the experimental
observations, indeed appear and are indicated by the black arrows
in the graphs. After invoking Snell's law, the negative refraction
results are deduced. From (a1) and (b1), we obtain the negative
refractive indices as $-0.84\pm 0.17$ and $-0.87\pm 0.21$ for the
two prisms respectively. The inconsistency between the two values
is less than 4\%, which is close to that estimated in the
experiment \cite{Chuang}. The overall uncertainty in the present
simulation (up to 24\%) is less than that in the experiment (up to
36\%) \cite{Chuang}. Therefore, a consistent negative refractive
index may be claimed from the measurements shown in
Fig.~\ref{fig1} (a1) and (b1). This would have been regarded as an
evidence showing that the photonic structures described by
Fig.~\ref{fig1} are left-handed materials, at least for the
frequency considered. By the same token, we even found that with
certain adjustments such as rotating the arrays or varying the
filling factor, thus obtained index can be close to the perfect
-1.

Though tempting, there are ambiguities in the above explanation of
the transmission in the context of negative refraction that has
led to the negative refraction index. This can be discerned by
taking into consideration the wave propagation inside the prisms.
As shown clearly by Fig.~\ref{fig1} (a2) and (b2), the
transmission inside the prisms has already been bent at the
incidence interfaces, referring to the two white arrowed lines in
(a2) and (b2). If Snell's law were valid at the outgoing
interfaces, it should also be applicable at the incident
boundaries. Then with the zero incidence angle and a finite
refraction angle, Snell's law would lead to the absurd result of
an infinite refractive index for the surrounding medium which is
taken as air. Besides, when taking into account the bending inside
the prism, the incident angle at the outgoing or the tilted
surface is not $\theta$ any more. Therefore the index value
obtained above cannot be correct. These ambiguities cannot be
excluded from the current experiments that have been claimed to
support LHM using Snell's law \cite{Chuang,Parazzoli}. In the
experiments of Ref.~\cite{Chuang}, for example, the transmission
fields inside the prisms are not shown. The authors also stated
clearly that the negative index behavior can be only achieved by
some adjustments, implying that the experimentally observed
negative index behavior is not robust. We have checked this point
by rotating the orientation of the arrays of the cylinders. The
apparent `negative refraction' may disappear, depending on the
relative geometries between the incident direction and the
orientation of the lattice.

In search for the cause of the apparent `negative refraction'
shown in Fig.~\ref{fig1}, we have carried out further simulations.
For example, in Fig.~\ref{fig2}, we plot the EM wave transmission
across two slabs of photonic crystals with two different lattice
orientations: one is along the diagonal direction, i.~e. the [11]
direction, and the other is along the direction making an equal
angle to the [11] and [10] directions. The incident frequency is
again 0.192 $*2\pi c/a$. Here, we see that when the incident wave
is along the [11] direction, the transmission follows a straight
path inside the slab. For the second case, the propagation
direction is bent inside the slab: the major lobe tends to follow
the [11] direction, and a minor lobe of transmission appears to
make an perpendicular angle with regard to the [11] direction,
that is the [-11] direction. As aforementioned, if Snell's law
were used to obtain the refraction index, the absurd number of
infinity would be deduced only ambiguously.

Fig.~\ref{fig2} clearly indicates that there are some favorable
directions for waves to travel. This phenomenon is not uncommon
for wave propagation in periodically structured media. It is
well-known that periodic structures can modulate dramatically the
wave propagation due to Bragg scattering, a feature fundamental to
the X-ray imaging. In some situations, the systems reveal complete
band gaps, referring to the frequency regime where waves cannot
propagate in any direction. In some other occasions, partial band
gaps may appear so that waves may be allowed to travel in some
directions but not in some other directions. More often, the
periodic structures lead to anisotropic dispersions in wave
propagation. The whole realm of these phenomena has been well
represented by the band structures calculated from Bloch's
theorem. The preferable directions for wave propagation are
associated with the properties of band structures. In fact, the
above abnormal `negative refraction' phenomenon may be explained
in the frameworks of the band structures calculated for the
crystal structures considered.

Fig.~\ref{fig3} shows the band structure calculated for the square
lattices used in the above simulations. A complete band gap is
shown between frequencies of 0.22 and 0.28. Just below the
complete gap, there is a regime of the partial band gap, in which
waves are not allowed to travel along the [10] direction. The
frequency 0.192$*2\pi c/a$ we have chosen for simulation lies
within this partial gap area. Within this gap, waves are
prohibited from transmission along the [10] direction. As a
result, when incident along an angle that lies between the [11]
and [10] direction, waves will incline to the [11] direction. This
observation explains the abnormal refraction observed above. Due
to the symmetry, the waves are also allowed to travel in the
directions which are perpendicular to [11]. This explains why we
have also seen an intensity lobe along [-11] from Fig.~\ref{fig2}.
We found that these results are also qualitatively similar for
other frequencies within the partial band gap. Furthermore, we
have checked other frequencies at which although there is no
partial band gap, the dispersion is highly anisotropic, similar
features may also be possible.

The phenomenon of the band structures may help discuss the
abnormal transmission that has been interpreted as the onset of
the negative index behavior in the experiments \cite{Chuang}. Up
to date, all the claimed left-handed materials are artificially
made periodic composite structures, in which the negative
permittivity and permeability are obtained near a resonance
frequency. From our experiences, near such a resonance frequency,
either complete band gaps or partial band gaps are likely to
appear. In other words, the wave propagation largely depends on
the periodic structures. Without taking this fact into account,
the transmission may be regarded as unconventional, and then the
artifact conclusion that a LHM has been realized and observed may
be resorted to. Due to the limited information available from the
experiments, we cannot yet definitely come to the conclusion that
the apparent `negative refractions' observed in experiments are
due to partial band gaps, highly anisotropic dispersions, or
inhomogeneities. The present simulations, however, at least
indicate that some of these effects must be addressed in order to
be able to interpret the experimental data correctly.

Lastly, we would like to make a comment on the permittivity and
permeability of composite materials. It is well-known that these
quantities are only meaningful in the sense of the effective
medium theory. A basic assumption is that the wave propagation
should not be sensitive to details of the supporting media.
Otherwise, whether the definitions of the permittivity and
permeability are still meaningful is itself doubtful. If
measurements of the two parameters could not be justified
accordingly, it could be misleading to deliberately interpret wave
propagation in terms of negative or positive refraction. In the
above cases, although the wave length is much larger than the
lattice constant, the wave propagation is obviously still
sensitive to the lattice arrangements. Therefore, the flow of
energy should be interpreted as mainly controlled by Bragg
scattering processes, rather than as a propagation in an effective
medium. In other words, the media act as a device that can bend
the flow of optical energies, rather than as an effective
refracting-medium.

In summary, we have simulated wave transmission through prism
structures. The results have shown some ambiguities in
interpreting the apparent abnormal behavior in the transmission as
the onset of the negative refraction index behavior. It is
suggested that periodic structures can give rise to peculiar
phenomena which need not be regarded as a negative index behavior.

{\bf Acknowledgments} The work received support from NSC and NCU.

\newpage

\begin{figure}[hbt]
\begin{center}
\caption{ \label{fig1}\small
The images of the transmitted intensity fields. Here, the
intensities inside and outside the prism structures are plotted
separately, so that the images can be clearly shown with proper
scales. The geometric measurements can be inferred from the
figure. The tilt angles for the two prisms are 18$^o$ and 26$^o$
respectively, and have been labelled in the figures. For cases
(a1) and (b1), we observe the apparent `negative refraction' at
the angles of 15$\pm 3^o$ and $21\pm 5^o$ respectively. When
applying Snell's law, these numbers give rise to the negative
refraction indices of $-0.84 \pm 0.17$ and $-0.87\pm 0.21$ for
(a1) and (b1) separately. In (a2) and (b2), the intensity fields
inside inside the prisms are plotted. Here we clearly see that the
transmission has been bent. In the plots, [10] and [11] denote the
axial directions of the square lattice of the cylinder
arrays.}\end{center}
\end{figure}

\begin{figure}[hbt]
\begin{center}
\caption{ \label{fig2}\small
The imaging fields for slabs of photonic crystal structure. Two
lattice orientations are considered: (a) The slab measures as
about 56$\times 10$ and the incidence is along the [11] direction;
(2) the slab measures as 50$\times 13$ and the incidence is along
the direction that makes an equal angle to the [11] and [10]
directions. The main lobes in the transmitted intensities are
shown.}
\end{center}
\end{figure}

\begin{figure}[hbt]
\begin{center}
\caption{ \label{fig3}\small
The band structure a square lattice of dielectric cylinders. The
lattice constant is $a$ and the radius of the cylinders is $0.3a$.
The $\Gamma M$ and $\Gamma X$ denote the [11] and [10] directions
respectively. A partial gap is between the two horizontal lines.}
\end{center}
\end{figure}

\end{document}